\documentclass[a4paper,10pt,onecolumn]{article}
\usepackage{amsmath,amsfonts}
\usepackage[dvips]{graphicx}
\begin{document}
\title{\bf Brane $f(\cal R)$ gravity}
\author{Ahmad Borzou\footnote{Electronic address: a.borzou@mail.sbu.ac.ir} ,\quad
Hamid Reza Sepangi\footnote{Electronic address:
hr-sepangi@sbu.ac.ir} ,\quad Shahab Shahidi\footnote{Electronic
address: sh.shahidi@mail.sbu.ac.ir}\quad and\quad Razieh
Yousefi\footnote{Electronic address: raz.yousefi@mail.sbu.ac.ir}
\\ {\small  Department of Physics, Shahid Beheshti University, Evin, Tehran 19839, Iran}}
\maketitle
\begin{abstract}
We consider a brane world scenario in which the bulk action is
assumed to have the form of a generic function of the Ricci scalar
$f(\mathcal{R})$ and derive the resulting Einstein field equations
on the brane. In a constant curvature bulk a conserved geometric
quantity appears in the field equations which can be associated with
matter. We present cosmological and spherically symmetric solutions
by assuming specific forms for $f(\mathcal{R})$ and show that the
former can explain an accelerated expanding universe while the
latter may account for galaxy rotation curves.
\end{abstract}
\section{Introduction}
The roots of the idea that our four-dimensional universe might be a
three-brane embedded in a higher dimensional bulk can be traced back
to string theory \cite{1}. In fact, the present day brane world
scenarios in their various flavors are influenced, in one way or
other, by string theory. One of the most successful of such higher
dimensional models is that proposed by Randall and Sundrum where our
4-dimensional world is considered as a brane in a 5-dimensional bulk
having the geometry of an AdS space \cite{2,3}. They were successful
in explaining what is known as the hierarchy problem; the enormous
disparity between the strength of the fundamental forces. The
Randall-Sundrum  (RS) scenario has had a great impact on our
understanding of the universe and has brought higher dimensional
gravitational theories to the fore. In the RS type models, only
gravitons can wander into the bulk whereas ordinary matter and gauge
fields are confined to the brane. Such a confinement is achieved by
the application of the Israel junction conditions. Later on,
Shiromizu, Maeda and Sasaki (SMS) \cite{5} showed how to
systematically project the Einstein field equations, written in the
bulk, onto the brane using the Gauss-Codazzi equations and the
Israel junction conditions \cite{4}, assuming $\mathbb{Z}_2$
symmetry. These field equations differ from the standard Einstein
field equations in $4D$ in that they have additional terms like
$\pi_{\mu\nu}$ which depend on the energy-momentum tensor on the
brane and the electric part of the Weyl tensor $E_{\mu\nu}$, leading
to the appearance of a quadratic term in the friedmann equation
\cite{6}.  Brane world scenarios have paved the way for a new
understanding of dark matter among other things for which static,
spherically symmetric solutions of the field equations on the brane
are of primary interest \cite{6a}. As an example, the galaxy
rotation curves have been accounted for in \cite{7} by studying the
vacuum solutions in a brane world scenario in which the brane admits
a family of conformal Killing vector fields, leading to exact
solutions of the field equations in parametric forms. These
solutions were then used to explain the galaxy rotation curves
without postulating the existence of dark matter.

Other brane world scenarios have emerged since the seminal work of
RS, the most influential of which is the model proposed by
Dvali-Gabadadze-Porrati (DGP) \cite{9}. In this model the bulk space
is assumed to be flat but there is an additional $4D$ induced
gravity term appearing in the action. The cosmological implications
of the DGP model was first studied in \cite{10} where it was shown
that the universe experiences a self accelerating phase at late
times. The DGP model predicts modifications to gravity at large
distances, in contrast to RS type models which modify gravity at
small scales. Other brane world models have been proposed in which
the condition of $\mathbb{Z}_2$ symmetry is relaxed and the
confinement of matter to the brane is achieved through the use of a
confining potential  \cite{11, 12, 13}. Brane world models in the
context of higher order gravity have also been studied
\cite{14,15,16} where the Gauss-Bonnet term is included in the bulk
action and  the $4D$ field equations are derived in the usual
manner.

Theories of gravity in which the Einstein-Hilbert (EH) action is
replaced with a generic function of $\cal R$ have been gaining
momentum in the last few years. In this regard, brane world
scenarios have also been proposed both in the context of $5D$
scalar-tensor type models \cite{19} where such a function of the
Ricci scalar is assumed to form the action in the bulk and DGP
models \cite{20} where the induced action on the brane contains the
$f({\cal R})$ term. However, the study of a RS type model with
$f({\cal R})$ as the action in bulk space has been lacking. Such a
setup would be of interest since there is no a priori reason as to
why the action in $5D$ should be the standard EH, in contrast to
$4D$ where the standard EH action seems to be, to a large extent,
adequate in describing the cosmos. It is therefore the purpose of
the present work to obtain the Einstein field equations on the brane
from a $f({\cal R})$ action in the bulk within the framework of the
RS model, using the SMS procedure. In doing so, we obtain the term
$Q_{\mu\nu}$ which completely depends on $f(\mathcal{R})$ and its
derivatives. Interestingly, for a conformally flat bulk space, this
quantity is conserved, that is, $\nabla^\mu Q_{\mu\nu}=0$. This is
of course welcome since one may identify such a term with some kind
of matter whose origins lie in the geometry of the bulk. It should
also be mentioned that the work presented in \cite{c1} is an attempt
to assume a generic $f({\cal R})$ function as the action for the
bulk space in the presence of a real scalar field. However, their
method of obtaining the field equations on the brane is based on
assuming a specific metric for the bulk which makes dealing with the
problem different from what we propose to do.

In what follows, we first derive the field equations on the brane
from a generic $f(\mathcal{R})$ action assumed for the bulk. We then
study both the static, spherically symmetric and cosmological
solutions of the resulting field equations. Conclusions are drawn in
the last section.

\section{Effective field equations on the brane}
We start by deriving the Einstein field equations on the brane,
assuming the following action for the bulk
\begin{align}
S=\int\mathrm{d}^5x\sqrt{-\mathrm{g}}~[f(\mathcal{R})+\mathcal{L}_m],\label{eq1}
\end{align}
where $\mathcal{L}_m$ is the matter lagrangian, $\mathrm{g}$ is the
bulk metric and $\mathcal{R}$ is the bulk Ricci scalar. After
variation of $S$ with respect to the bulk metric $\mathrm{g}_{AB}$,
we obtain
\begin{align}
f^\prime(\mathcal{R})\mathcal{R}_{AB}&-\frac 1 2 \mathrm{g}_{AB}f(\mathcal{R})+\mathrm{g}_{AB}\Box f^\prime(\mathcal{R})\nonumber \\
&-\nabla_A\nabla_B f^\prime(\mathcal{R})=\kappa_5^2 T_{AB},\label{eq2}
\end{align}
where a prime represents derivative with respect to the argument.
Rearranging the above equation, we obtain the effective Einstein
field equations in the bulk
\begin{align}
G_{AB}\equiv\mathcal{R}_{AB}-\frac 1 2 \mathcal{R}\mathrm{g}_{AB}=T^{tot}_{AB},\label{eq3}
\end{align}
where
\begin{align}
T^{tot}_{AB}&=\frac{1}{f^\prime(\mathcal{R})}\bigg[\kappa_5^2 T_{AB}-\Big(\frac 1 2 \mathcal{R}f^\prime(\mathcal{R})
-\frac 1 2 f(\mathcal{R})\nonumber \\
&+\Box f^\prime(\mathcal{R})\Big)\mathrm{g}_{AB}+\nabla_A\nabla_B f^\prime(\mathcal{R})\bigg].\label{eq4}
\end{align}
To obtain the field equations on the brane we start by invoking the
Gauss equations \cite{23}
\begin{align}
R^{\alpha}_{~\beta\gamma\delta}=\mathcal{R}^{A}_{~BCD}h^{\alpha}_{~A} h^B_{~\beta}h^C_{~\gamma}h^D_{~\delta}+
K^{\alpha}_{~\gamma}K_{\beta\delta}-K^{\alpha}_{~\delta}K_{\beta\gamma},\label{eq5}
\end{align}
where $h_{AB}=\mathrm{g}_{AB}-n_{A}n_{B}$ is the induced metric on
the brane,  $n^{A}$ is the unit vector normal to the
four-dimensional brane, $R^{\alpha}_{~\beta\gamma\delta}$ is the
Riemann tensor on the brane and $K_{\mu\nu}=h_{~\mu}^A
h_{~\nu}^B\nabla_A n_B$ is the extrinsic curvature. After
contracting equation (\ref{eq5}) one obtains
\begin{align}
R_{\mu\nu}&=\mathcal{R}_{AB} h^A_{~\mu} h^B_{~\nu}-\mathcal{R}^A_{~BCD} n_{A}h^B_{~\mu}n^C h^D_{~\nu} \nonumber \\
&+K K_{\mu\nu} -K^{\alpha}_{~\mu}K_{\nu\alpha},\label{eq6}
\end{align}
and
\begin{align}
G_{\mu\nu}&=\left[\mathcal{R}_{AB}-\frac 1 2 \mathrm{g}_{AB}
\mathcal{R}\right]h^A_{~\mu}h^B_{~\nu}+\mathcal{R}_{AB}n^A n^B h_{\mu\nu}\nonumber\\&+ K K_{\mu\nu}-K^{\rho}_{~\mu}K_{\nu\rho}-
\frac 1 2 h_{\mu\nu}(K^2-K^{\alpha\beta}K_{\alpha\beta})\nonumber \\
&-\mathcal{R}^A_{~BCD}n_A n^C h^B_{~\mu} h^D_{~\nu}.\label{eq7}
\end{align}
Now, let us take the usual decomposition of the Riemann tensor into
the Ricci tensor, Ricci scalar and the Weyl tensor
$\mathcal{C}_{ABCD}$ as \cite{5}
\begin{align}
\mathcal{R}_{ABCD}&=\frac 2 3
(\mathrm{g}_{A[C}{\mathcal{R}}_{D]B}-\mathrm{g}_{B[C}{\mathcal{R}}_{D]A})\nonumber \\
&-\frac{1}{6}\mathrm{g}
_{A[C}\mathrm{g}_{D] B}{\mathcal{R}}+ \mathcal{C}_{ABCD},\label{eq8}
\end{align}
and using equation (\ref{eq3}) to obtain
\begin{align}
&G_{\mu\nu}=\frac{2}{3}\left[T^{tot}_{AB}h^A_{~\mu}h^B_{~\nu}+
\left(T^{tot}_{AB}n^{A}n^{B}-\frac{1}{4}T^{tot}\right) h_{\mu\nu}\right]\nonumber \\
&+KK_{\mu\nu}-K^{\sigma}_{~\mu}K_{\nu\sigma}-\frac{1}{2}h_{\mu\nu}(K^2-K^{\alpha\beta}K_{\alpha\beta})-E_{\mu\nu},\label{eq9}
\end{align}
where $T^{tot}$ is the trace of $T^{tot}_{AB}$ and
\begin{align}
E_{\mu\nu}=\mathcal{C}^A_{~BCD}n_A n^C h^B_{~\mu}h^D_{~\nu},\label{eq10}
\end{align}
is the electric part of the Weyl tensor. If we assume that the brane
is located at $y=0$  we can write the energy-momentum tensor as
\cite{5}
\begin{align}
T_{AB}=-\Lambda g_{AB}+S_{AB}\delta(y),\label{eq11}
\end{align}
where $\Lambda$ is the cosmological constant of the bulk and
\begin{align}
S_{\mu\nu}=-\lambda h_{\mu\nu}+\tau_{\mu\nu},\label{eq12}
\end{align}
with $\tau_{\mu\nu}$ being the brane energy-momentum tensor and
$\lambda$ the associated cosmological constant. With the use of
equation (\ref{eq11}) we can compute the first three terms in
equation (\ref{eq9}) as
\begin{align}
T^{tot}_{AB}h^A_{~\mu}h^{B}_{~\nu}&=
\left[-\frac{1}{f^\prime(\mathcal{R})}\kappa_5^2\Lambda-\frac 1 2 \mathcal{R}+\frac 1 2\frac{f(\mathcal{R})}
{f^\prime(\mathcal{R})}-\frac{\Box f^\prime(\mathcal{R})}{f^\prime(\mathcal{R})}\right]h_{\mu\nu}\nonumber \\
&+\left[\frac{\nabla_A\nabla_B f^\prime(\mathcal{R})}{f^\prime(\mathcal{R})}\right]h^A_{~\mu}h^B_{~\nu},\label{eq13}
\end{align}
\begin{align}
T^{tot}_{AB}n^{A}n^{B}h_{\mu\nu}&=
\left[-\frac{1}{f^\prime(\mathcal{R})}\kappa_5^2\Lambda-\frac 1 2
\mathcal{R}+\frac 1 2\frac{f(\mathcal{R})}
{f^\prime(\mathcal{R})}-\frac{\Box
f^\prime(\mathcal{R})}{f^\prime(\mathcal{R})}\right]h_{\mu\nu}\nonumber\\&+\left[
\frac{\nabla_A\nabla_B
f^\prime(\mathcal{R})}{f^\prime(\mathcal{R})}\right]n^{A}n^{B}h_{\mu\nu},\label{eq14}
\end{align}
\begin{align}
-\frac{1}{4}T^{tot}h_{\mu\nu}=
\left(\frac 5 4 \frac{1}{f^\prime(\mathcal{R})}\kappa_5^2\Lambda+\frac 5 8 \mathcal{R}-\frac 5 8
\frac{f(\mathcal{R})}{f^\prime(\mathcal{R})}+\frac{\Box f^\prime(\mathcal{R})}{f^\prime(\mathcal{R})}\right)h_{\mu\nu}.\label{15}
\end{align}
To compute the last terms in equation (\ref{eq9}) we use the Israel
junction conditions \cite{4,5}
\begin{align}
[K_{\mu\nu}]=-\kappa_{5}^2\left(S_{\mu\nu}-\frac{1}{3}h_{\mu\nu}S\right),\label{eq16}
\end{align}
where $[X]:=\lim_{y\rightarrow 0^+}X-\lim_{y\rightarrow 0^-}X$ is
the jump of the quantity $X$ across the brane. Assuming
$\mathbb{Z}_{2}$ symmetry, we can replace the jump in the extrinsic
curvature by the value of the extrinsic curvature at the location of
the brane, hence
\begin{align}
K^+_{\mu\nu}=K^-_{\mu\nu}=-\frac{\kappa_{5}^2}{2}\left(S_{\mu\nu}-\frac{1}{3}h_{\mu\nu}S\right),\label{eq17}
\end{align}
Now, by expressing the extrinsic curvature in terms of
$\tau_{\mu\nu}$ and $\lambda$ and using the relation
\begin{align}
\mathcal{R}=-\frac{5}{f^\prime(\mathcal{R})}\kappa_5^2\Lambda+\frac
5 2 \frac{f(\mathcal{R})} {f^\prime(\mathcal{R})}-4 \frac{\Box
f^\prime(\mathcal{R})}{f^\prime(\mathcal{R})},\label{eq19}
\end{align}
obtained by contracting equations(\ref{eq2}), we finally obtain the effective Einstein equations
on the brane
\begin{align}
G_{\mu\nu}=-\Lambda_4 h_{\mu\nu}+8\pi
G_N\tau_{\mu\nu}+\kappa_{5}^{2}\pi_{\mu\nu}+Q_{\mu\nu}-E_{\mu\nu},\label{eq20}
\end{align}
where
\begin{align}
\Lambda_4=\frac 1 2 \kappa_5^2\left(\Lambda+\frac 1 6
\kappa^2_5\lambda^2\right),\label{eq21}
\end{align}
\begin{align}
G_{N}=\frac{\kappa_{5}^4\lambda}{48\pi},\label{eq22}
\end{align}
\begin{align}
\pi_{\mu\nu}=-\frac{1}{4}\tau_{\mu\alpha}\tau_{\nu}^{\alpha}+\frac{1}{12}\tau\tau_{\mu\nu}+\frac{1}{8}
h_{\mu\nu}\tau_{\alpha\beta}\tau^{\alpha\beta}-
\frac{1}{24}h_{\mu\nu}\tau^2,\label{eq23}
\end{align}
\begin{align}
Q_{\mu\nu}=\left[F(\mathcal{R})h_{\mu\nu}+\frac 2 3
\frac{\nabla_A\nabla_B f^\prime(\mathcal{R})}{f^\prime(\mathcal{R})}
\left(h^A_{~\mu}h^B_{~\nu}+n^{A}n^{B}h_{\mu\nu}\right)\right]_{y=0},\label{eq24}
\end{align}
and
\begin{align}
F(\mathcal{R})\equiv &-\frac{4}{15}\frac{\Box
f^\prime(\mathcal{R})}{f^\prime(\mathcal{R})}-\frac{1}{10}\mathcal{R}
\left(\frac 3 2 +f^\prime(\mathcal{R})\right) \nonumber \\
&+\frac 1 4 f(\mathcal{R})-\frac 2 5 \Box f^\prime(\mathcal{R}).\label{eq25}
\end{align}
We note that for $f({\cal R})={\cal R}$, we obtain the usual induced
equations first obtained in \cite{5}.  We also note that in the case
of constant curvature bulk we have $E_{\mu\nu}=0$ and by equation
(\ref{eq24}) and (\ref{eq25}) we obtain
\begin{align}
\nabla^\mu Q_{\mu\nu}=0,\label{eq26}
\end{align}
since $f(\mathcal{R})$ is a constant. This means that we can
identify $Q_{\mu\nu}$ as representing the energy-momentum tensor of
an unknown type of matter. In the case of an arbitrary curvature
bulk we have the following conservation equation
\begin{align}
\nabla^\mu\left(Q_{\mu\nu}-E_{\mu\nu}\right)=0.\label{eq27}
\end{align}
\section{$f(\mathcal{R})=\mathcal{R}^n$ solutions}
In this section we consider static, spherically symmetric and
cosmological solutions to the Einstein field equations given by
equation (\ref{eq20}). In what follows, we start by studying the
black hole solutions and then move on to study the cosmological
solutions. In all the solutions presented below, we assume a
specific form for the function $f({\cal R})$, namely
$f(\mathcal{R})=\mathcal{R}^n$.
\subsection{Spherically symmetric solutions}
We begin by taking the following static, spherically symmetric bulk
metric
\begin{align}
ds^{2}=-A(r)dt^{2}+\frac{1}{A(r)}dr^{2}+r^2d\theta^{2}+r^2\sin\theta^2d\phi^2+dy^2.\label{eq28}
\end{align}
If we assume that the brane is devoid of matter then the second and
third terms in (\ref{eq20}) become zero. Taking the normal vector to
the brane as $n^A=(0,0,0,0,1)$, the Einstein field equation becomes
\begin{align}
G_{\mu\nu}=-\Lambda_4 h_{\mu\nu}+Q_{\mu\nu}-E_{\mu\nu},\label{eq29}
\end{align}
where $Q_{\mu\nu}$ is given by equation (\ref{eq24}). Let us take
the constant curvature solutions of equation (\ref{eq29}) as
\begin{align}
A(r)=1+\frac a r + br^2\label{eq30}.
\end{align}
The components of $Q^\mu_{~\nu}$ are then given by
\begin{align}
  Q^0_{~0}=Q^1_{~1}=Q^2_{~2}=Q^3_{~3}=\frac{9b}{5}-\frac{n(-12b)^n}{10}+\frac{(-12b)^n}{4}, \label{eq31}
\end{align}
where $b$ is a
constant depending on the cosmological constant $\Lambda_4$ and
$n$ and obtained from the equation
\begin{align}
&\left(\frac{n}{10}-\frac{1}{4}\right)(-12b)^n+\frac{6b}{5}+\Lambda_4=0,\qquad n\neq 1, \label{eq32}\\
&b=-\frac {\Lambda_4} {3},\qquad n=1. \label{eq33}
\end{align}
Equation (\ref{eq30}) represents a Schwarzschild-Anti de-Sitter like
solution which is compatible with \cite{26}, so we can identify $a$
by $-2M$ as in ordinary schwarzschild solutions. In the case $n=1$
we recover the standard Schwarzschild-Anti de-Sitter solution. In
the case of a vanishing brane cosmological constant, we also have
the above solution reduced to
\begin{align}
b=-\frac{1}{12}\left(\frac{2}{2n-5}\right)^\frac{1}{n-1},\qquad
n\neq 1,\label{eq34}
\end{align}
depending only on $n$. Therefore, $Q_{\mu\nu}$ plays the role of the
cosmological constant in this case. The form of the above solution
suggests that, in very general terms, it may be used to explain the
galaxy rotation curves.

For the sake of completeness, let us also consider the case where
$f(\mathcal{R})=\mathcal{R}+\mathcal{R}^n$ which has been considered
in many works. In this case, the equation for $B(n)$ is similar to
equation (\ref{eq32}) except  that the coefficient  $b$ is replaced
by $3$. In the case of vanishing cosmological constant we find
\begin{align}
b=-\frac{1}{12}\left(\frac{5}{2n-5}\right)^\frac{1}{n-1},\qquad
n\neq 1.\label{eq34.5}
\end{align}
\subsection{Cosmological solutions}
In order to investigate the cosmological solutions we consider the
standard Robertson-Walker metric
\begin{align}
ds^2 = -dt^2+a(t)^2\left[dr^2+r^2 d\theta^2+r^2 \sin^2\theta
d\phi^2\right]+dy^2,\label{eq35}
\end{align}
where $a(t)$ is the scale factor. We now consider a perfect fluid
form for the brane matter with a linear equation of state
\begin{align}
&\tau^\mu_{~\nu}=\mbox{diag}(-\rho(t),p(t),p(t),p(t)),\label{eq35}\\
&p(t)=\omega\rho(t),\label{eq37}
\end{align}
where $\omega$ is a constant. Again, with $n^A=(0,0,0,0,1)$ and the
assumption that the cosmological constants in both the bulk and
brane are zero, we obtain the Einstein field equations as follows
\begin{align}
G_{\mu\nu}=\kappa_{5}^{2}\pi_{\mu\nu}+Q_{\mu\nu}-E_{\mu\nu}.\label{eq38}
\end{align}
Let us consider a power law form for the scale factor
\begin{align}
a(t)=\left(\frac{t}{t_0}\right)^B.\label{eq39}
\end{align}
The Einstein equations then become
\begin{align}
A_1+c\left(\frac{6B(2B-1)}{t^2}\right)^n t^2=5\kappa_{5}^{2} a_1
t^2\rho(t)^2,\label{eq41}
\end{align}
\begin{align}
A_2+c\left(\frac{6B(2B-1)}{t^2}\right)^n t^2=5\kappa_{5}^{2} a_2
t^2\rho(t)^2,\label{eq42}
\end{align}
where
\begin{align}
&A_1=24B(2B-1)(3B+2n-1)(B-2n+2),\label{eq43}\\
&A_2=8B(2B-1)(9B^2-5B-2nB-12n+8n^2+4),\label{eq44}\\
&c=(30-12n)B^2+8n(1-3n+2n^2)-3B(8n^2-10n+5),\label{eq45}\\
&a_1=B(2B-1),\label{eq46}\\
&a_2=-B(2B-1)(2\omega+1).\label{eq47}
\end{align}
The solution is given by
\begin{align}
\omega=-1\quad \mbox{for}~B\neq 0,\qquad
B(n)=\frac{1-3n+2n^2}{2-n}.\label{eq48}
\end{align}
\begin{figure}
  \centering
  \includegraphics[scale=0.5]{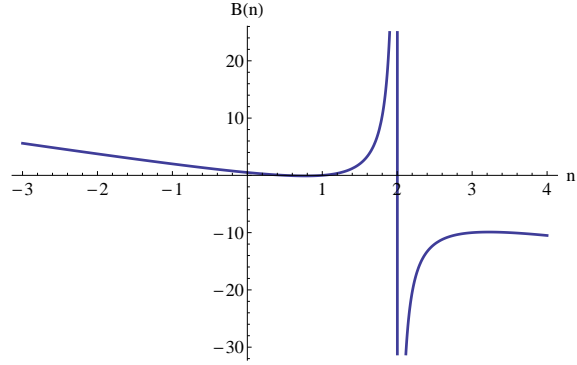}
  \caption{plot of $B(n)$ as a function of $n$.}
  \label{fig1}
\end{figure}
In figure \ref{fig1} we have plotted $B(n)$ as a function of $n$.
For an accelerating universe we must have $B>1$. Now, using equation
(\ref{eq41}) we obtain
\begin{align}
\rho(t)^2=\frac{1}{5\kappa_{5}^{2}
a_1}\left[\frac{A_1}{t^2}+c\left(\frac{6B(2B-1)}{t^2}\right)^n\right],\label{eq49}
\end{align}
where
\begin{align}
&A_1=A_2=\frac{24n(2n-1)^3(n-1)^2(4n-5)^2}{(n-2)^4},\label{eq50}\\
&c=-\frac{n(n-1)(2n-1)^2(4n-5)}{(n-2)^2},\label{eq51}\\
&a_1=a_2=B(B-1).\label{eq52}
\end{align}
\begin{figure}
  \centering
  \includegraphics[scale=0.5]{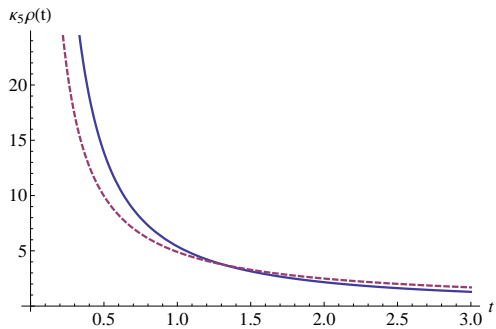}
  \caption{plot of $\rho(t)$ as a function of $t$ for $f(\mathcal{R})=\mathcal{R}+\mathcal{R}^n$,
  dotted curve and for
  $f(\mathcal{R})=\mathcal{R}^n$, solid curve, both for $n=\sqrt{2}$.}
  \label{fig2}
\end{figure}
In the case $f(\mathcal{R})=\mathcal{R}+\mathcal{R}^n$ we have the solution
\begin{align}
\omega=-1~ \mbox{for}~B\neq 0,\qquad
B(n)=\frac{7n-4\pm\sqrt{n(4n-1)(8n^2-7n+2)}}{8n^2+3n-8}.
\end{align}
Figure \ref{fig2} shows a plot  of $\rho(t)$ as a function of $t$
for $n=\sqrt{2}$ in both  $f(\mathcal{R})=\mathcal{R}^n$ and
$f(\mathcal{R})=\mathcal{R}+\mathcal{R}^n$ cases, taking the plus
sign in $B(n)$ above. Using equation (\ref{eq48}) we find $B=1.47$
which would point to an accelerated expanding universe.
\section{Conclusions}
In this paper we have studied a brane world scenario in which the
bulk action consists of a generic function of the Ricci scalar
$f({\cal R})$.  The field equations on the brane were then obtained
using the SMS method. In this process we obtained the quantity
$Q_{\mu\nu}$ which originates from  the geometry of the bulk space.
For a conformally flat bulk, this quantity is conserved and could
therefore be identified with a new kind of matter. We also
investigated the static and cosmological solutions of the field
equations on the brane and found that the former could be used to
explain the galaxy rotation curves while the latter gives the
observed behavior for the scale factor at late times.
{}

\begin{thebibliography}{99}
\bibitem{1} M. Green, J. Schwarz, and E. Witten, \textit{Superstring Theory}
(Cambridge Monographs on Mathematical Physics) two volumes,
Cambridge University Press, 1987; Lust and Theisen,
\textit{Lectures on String Theory} Springer Verlag, 1989; J.
Polchinski, \textit{String Theory} Cambridge Monographs on
Mathematical Physics, two volumes, Cambridge University Press,
2004; B. Zweibach, \textit{A First Course in String Theory}
Cambridge University Press, 2004.
\bibitem{2} L. Randall and R. Sundrum, Phys. Rev. Lett. 83 (1999)
3370 [arXiv:hep-ph/9905221]
\bibitem{3} L. Randall and R. Sundrum, Phys. Rev. Lett. 83 (1999) 4690
[arXiv:hep-th/9906064]
\bibitem{4} W. Israel, Nouvo Cimento B 44 (1966) 1 .
\bibitem{5} T. Shiromizu, K. Maeda, M. Sasaki, Phys. Rev. D 62 (2000)
024012 [arXiv:gr-qc/9910076]
\bibitem{6} P. Binetruy, C. Deffayet and D. Langlois, nucl. Phys. B 565 (2000) 269 [arXiv:hep-th/9905012];
P. Binetruy, C. Deffayet,  U. Ellwanger and D. Langlois, Phys. Lett.
B 477 (2000) 285 [arXiv:hep-th/9910219].
\bibitem{6a} N. Dadhich, R. Maartens, P. Papadopoulos and V.
Rezania, Phys. Lett. B 487 (2000) 1 [arXiv:hep-th/0003061].
\bibitem{7} T. Harko and M. K. Mak, Phys. Rev. D 69 (2004) 064020 [arXiv:gr-qc/0401049];
T. Harko and M. K. Mak, Phys. Rev. D 70 (2004) 024010
[arXiv:gr-qc/0404104].
\bibitem{9}  G. Dvali, G. Gabadadze and M. Porrati, Phys. Lett. B 485 (2000) 208 [arXiv:hep-th/0005016].
\bibitem{10} C. Deffayet, Phys. Lett. B 502 (2001) 199 [arXiv:hep-th/0010186].
\bibitem{11} S. Jalalzadeh and H. R. Sepangi, Class. Quant. Gravity 22 (2005) 2035 gr-qc/0408004 .
\bibitem{12} M. Heydari-Fard, H. Razmi and H. R. Sepangi, Phys. Rev. D 76 (2007) 066002 arXiv:0707.3558 .
\bibitem{13} M. Heydari-Fard, M. Shirazi, S. Jalalzade and H. R. Sepangi, Phys. Lett. B 640 (2006) 1 gr-qc/0607067 .
\bibitem{14} K. Maeda, T. Torii, Phys. Rev. D 69 (2004) 024002 hep-th/0309152 .
\bibitem{15} M. Heydari-Fard, H. R. Sepangi, Phys. Rev. D 75 (2007) 064010 [arXiv:gr-qc/0702061].
\bibitem{16} K. Konya, Class. Quant. Grav. 24 (2007) 2761 [arXiv:gr-qc/0605119].
\bibitem{19} K. Atazadeh, M. Farhoudi and H. R. Sepangi, Phys. Lett. B 660 (2008) 275 [arXiv:gr-qc/0801.1398v1].
\bibitem{c1}  V. I. Afonso, D. Bazeia, R. Menezes, A. Yu. Petrov, Phys. Lett. B 658 (2007) 71, [hep-th/0710.3790].
\bibitem{20} J. Saavedra and Y. Vasquez [arXiv: gr-qc/0803.1823v2];
{\it ibid} [arXiv:hep-th/0601213]; {\it ibid}
[arXiv:astro-ph/0801.4843].
\bibitem{23} L. P. Eisenhart, \textit{Riemannian geometry} (Princeton University Press, Princeton, NJ, 1966).
\bibitem{26} A. Chamblin, S. W. Hawking, H. S. Reall, Phys. Rev. D 61 (2000) 065007.
\end{thebibliography}
\end{document}